\documentclass{pasa}%

\usepackage{graphicx}
\usepackage{amsmath}	
\usepackage{amssymb}	
\usepackage{rotating}

\newcommand{\myaffil}[1]{$^{\rm #1}$}
\newcounter{inst}
\newcommand{\inst}[1]{\noindent%
   \refstepcounter{inst}\myaffil{\arabic{inst}\label{#1}}     
   }
   
\newcommand{\arcdeg}{\ensuremath{^{\circ}}}

\newcommand{\perbeam}{\,beam\ensuremath{^{-1}}}
\DeclareMathOperator{\arcsinh}{arcsinh}

\newcommand{\survarea}{2,860 }
\newcommand{\catarea}{2,670 }
\newcommand{\nfit}{17,244 }
\newcommand{\nsrc}{22,037 }
\newcommand{\nneg}{31 }
\newcommand{\nremoved}{60 }
\newcommand{\nvresolved}{168 }
\newcommand{\nresolved}{5,749 }

\newcommand{\ncol}{311 }
\newcommand{\pctreliable}{99.86}

\newcommand{\fig}{Fig.}

\newcommand{\Fig}{Fig.}
\newcommand{\Figs}{Figs.}
\newcommand{\sect}{Section}

\newcommand{\Sect}{Section}

\newcommand{\tab}{Table}

\newcommand{\Tab}{Table}

\newcommand{\farcs}{\mbox{\ensuremath{.\!\!^{\prime\prime}}}}
\newcommand{\fdg}{\mbox{\ensuremath{.\!\!^\circ}}}

\title[GLEAM Survey II: Galactic Plane]{GaLactic and Extragalactic All-sky Murchison Widefield Array (GLEAM) survey II: Galactic Plane $345^\circ < l < 67^\circ$, $180^\circ < l < 240^\circ$ }

\author[Hurley-Walker~et~al.]{N.~Hurley-Walker\myaffil{\ref{ICRAR}},
P.~J.~Hancock\myaffil{\ref{ICRAR},\ref{CAASTRO}},
T.~M.~O.~Franzen\myaffil{\ref{ASTRON}},
J.~R.~Callingham\myaffil{\ref{ASTRON}},
A.~R.~Offringa\myaffil{\ref{ASTRON}}, 
L.~Hindson\myaffil{\ref{Hert}},
C.~Wu\myaffil{\ref{UWA}},
M.~E.~Bell\myaffil{\ref{UTS}},
B.-Q.~For\myaffil{\ref{ASTRO3D},\ref{UWA}},
B.~M.~Gaensler\myaffil{\ref{ASTRO3D},\ref{Toronto}},
M.~Johnston-Hollitt\myaffil{\ref{ICRAR}},
A.~D.~Kapi\'nska\myaffil{\ref{NRAO}},
J.~Morgan\myaffil{\ref{ICRAR}},
T.~Murphy\myaffil{\ref{USyd},\ref{CAASTRO}},
B.~McKinley\myaffil{\ref{ICRAR}},
P.~Procopio\myaffil{\ref{CAASTRO},\ref{UMelb}},
L.~Staveley-Smith\myaffil{\ref{ASTRO3D},\ref{UWA}}
R.~B.~Wayth\myaffil{\ref{ICRAR},\ref{CAASTRO}},
Q.~Zheng\myaffil{\ref{SHAO}} \\
{\small \myaffil{}\,Email: nhw@icrar.org}\\
{\small \inst{ICRAR}\,International Centre for Radio Astronomy Research, Curtin University, Bentley, WA 6102, Australia}\\
{\small \inst{CAASTRO}\,ARC Centre of Excellence for All-sky Astrophysics (CAASTRO)}\\
{\small \inst{ASTRON}\,ASTRON, Netherlands Institute for Radio Astronomy, Oude Hoogeveensedijk 4, 7991 PD, Dwingeloo, The Netherlands}\\
{\small \inst{Hert}\,Centre for Astrophysics Research, School of Physics, Astronomy and Mathematics, University of Hertfordshire, College Lane, Hatfield AL10 9AB, UK}\\
{\small \inst{UWA}\,International Centre for Radio Astronomy Research, University of Western Australia, Crawley 6009, Australia}\\
{\small \inst{UTS}\,University of Technology Sydney, 15 Broadway, Ultimo NSW 2007, Australia}\\
{\small \inst{RRI}\,Raman Research Institute, Bangalore 560080, India}\\
{\small \inst{ASTRO3D}\,ARC  Centre  of  Excellence  for  All  Sky  Astrophysics  in  3  Dimensions  (ASTRO  3D)}\\
{\small \inst{Toronto}\,Dunlap Institute for Astronomy and Astrophysics, 50 St. George St, University of Toronto, ON M5S 3H4, Canada}\\
{\small \inst{NRAO}\,National Radio Astronomy Observatory, P.O. Box O, Socorro, NM 87801, USA}\\
{\small \inst{USyd}\,Sydney Institute for Astronomy, School of Physics, The University of Sydney, NSW 2006, Australia}\\
{\small \inst{UMelb}\,School of Physics, The University of Melbourne, Parkville, VIC 3010, Australia}\\
{\small \inst{SHAO}\,Shanghai Astronomical Observatory, 80 Nandan Rd, Xuhui Qu, Shanghai Shi, China, 200000}\\
}

\jid{PASA}
\doi{10.1017/pas.\the\year.xxx}
\jyear{\the\year}

\usepackage{aas_macros}
\usepackage{hyperref} 
\hypersetup{colorlinks,citecolor=blue,linkcolor=blue,urlcolor=blue}

\begin{document}

\begin{frontmatter}
\maketitle

\begin{abstract}
This work makes available a further \survarea~deg$^2$ of the GLEAM survey, covering half of the accessible Galactic Plane, across twenty frequency bands sampling 72--231\,MHz, with resolution 4'--2'.
Unlike previous GLEAM data releases, we used multi-scale \textsc{clean} to better deconvolve large-scale Galactic structure.
For the Galactic longitude ranges $345^\circ < l < 67^\circ$, $180^\circ < l < 240^\circ$, we provide a compact source catalogue of \nsrc components selected from a 60-MHz bandwidth image centred at 200-MHz, with RMS noise $\approx10$--20\,mJy\perbeam and position accuracy better than 2".
The catalogue has a completeness of 50\,\% at $\approx120$\,mJy, and a reliability of \pctreliable\,\%.
It covers Galactic latitudes $1^\circ\leq|b|\leq10^\circ$ toward the Galactic Centre and $|b|\leq10^\circ$ for other regions, and is available from Vizier; images covering $|b|\leq10^\circ$ for all longitudes are made available on the GLEAM VO server and SkyView.
\end{abstract}


\end{frontmatter}



\section{Introduction}

New low-frequency radio telescopes are exploring the sky at a rapid rate, driven by the goal of detecting the redshifted \textsc{Hi} signal of the Epoch of Reionisation. Foregrounds to this experiment are orders of magnitude larger than the signal and encompass emission from galaxies across the Universe, including our own Milky Way.

Radio emission in our own Galaxy is primarily from three main sources: the diffuse synchrotron emission from the interaction of the relativistic fraction of the interstellar medium (ISM) with Galactic magnetic fields; free-free emission from thermal \textsc{Hii} regions, and discrete synchrotron emitters like pulsar wind nebulae (PWNe), colliding-wind binary star systems, and supernova remnants (SNRs).

The Murchison Widefield Array \citep[MWA; ][]{2013PASA...30....7T}, operational since 2013, is a precursor to the low-frequency component of the Square Kilometre Array, which will be the world's most powerful radio telescope. The GaLactic and Extragalactic All-sky MWA \citep[GLEAM; ][]{2015PASA...32...25W} survey observed the whole sky south of declination (Dec) $+30^\circ$ from 2013 to 2015 between 72 and 231\,MHz. A major data release covering 24,402 square degrees of extragalactic sky was published by \cite{2017MNRAS.464.1146H}, while individual studies have published smaller regions such as the Magellanic Clouds \citep{2018MNRAS.tmp.1867F}.

The Galactic Plane poses a challenge to low-frequency imaging, as it produces large amounts of power on a range of spatial scales, and this power also changes with frequency. The spectral index of the diffuse emission is $\alpha=-0.7$, where the flux density $S$ at a frequency $\nu$ is given by $S_\nu \propto \nu^\alpha$. \textsc{Hii} regions typically have flatter spectral indices of $-0.2 < \alpha < +2$ \citep{2016era..book.....C} while SNR may have steeper spectral indices of $-1.1 < \alpha < 0$, depending on age and environment \citep[see][for a review]{2015A&ARv..23....3D}. The $(u,v)$-coverage of interferometric arrays also changes with frequency, adding a further difficulty to reconstructing an accurate image of these regions.

In this paper, we present the data reduction used to produce images covering the Galactic plane, and an associated catalogue, over the longitude range $345^\circ < l < 60^\circ$, $180^\circ < l < 240^\circ$, for latitudes $|b|<10^\circ$. \Sect~\ref{sec:dr} describes the observations and data reduction; \Sect~\ref{sec:images} presents the resulting images; \Sect~\ref{sec:catalogue} derives a compact source catalogue, \Sect~\ref{sec:galactic} discusses some of the subtleties of interpreting the Galactic plane images, and \Sect~\ref{sec:conclusions} concludes with thoughts on further work.


\section{Data reduction}\label{sec:dr}

Some common software packages are used throughout the data reduction. Unless otherwise specified:
\begin{itemize}
\item To convert radio interferometric visibilities into images, we use the widefield imager \textsc{WSClean} \citep{2014MNRAS.444..606O} version 2.3.4, which correctly handles the non-trivial $w$-terms of MWA snapshot images; versions 2 onward include useful features such as automatically-thresholded \textsc{clean}ing, and multi-scale \textsc{clean};
\item to mosaic together resulting images, we use the mosaicking software \textsc{swarp} \citep{2002ASPC..281..228B}; to minimise flux loss from resampling, images are oversampled by a factor of four when regridded, before being downsampled back to their original resolution;
\item to perform source-finding, we use \textsc{Aegean} v2.0.2\footnote{https://github.com/PaulHancock/Aegean} \citep{2012MNRAS.422.1812H, 2018PASA...35...11H} and its companion tools such as the Background and Noise Estimator (\textsc{BANE}); this package has been optimised for the wide-field images of the MWA, and includes the ``priorised'' fitting technique, which is necessary to obtain flux density measurements for sources over a wide bandwidth. Fitting errors calculated by \textsc{Aegean} take into account the correlated image noise, and are derived from the fit covariance matrix, which quantifies the quality of fitting; if the fit is poor, and the residuals are large, the fitting errors on position, shape, flux density etc all increase appropriately, so it produces useful error estimates for further use.
\end{itemize}

\subsection{Observations}

This paper covers the Galactic plane for two Galactic longitude ranges within galactic latitude $|b|\leq10^\circ$: $345^\circ < l < 67^\circ$ and $180^\circ < l < 240^\circ$, hereafter respectively referred to as inner-Galaxy (iG) and outer-Galaxy (oG) regions. The longitude range $240^\circ < l < 345^\circ$ is discussed in Johnston-Hollitt et al. (in prep). \cite{2017MNRAS.464.1146H} presented GLEAM observations of most of the extragalactic ($|b|>10^\circ$) sky, and during the data processing to produce those images, the oG region was imaged. Due to the paucity of bright diffuse emission in this region, that initial pipeline produced good-quality images which will be analysed later in the paper (\sect~\ref{sec:images}).

However, the iG region was not amenable to a pipeline optimised for imaging the extragalactic sky, due to an increased amount of power on larger scales. This region therefore required bespoke re-processing, which will be discussed here. The GLEAM survey strategy is described in detail by \cite{2015PASA...32...25W}, but we summarise it here. To cover 72--231\,MHz using the 30.72-MHz instantaneous bandwidth of the MWA, five frequency ranges of 72--103\,MHz, 103--134\,MHz, 139--170\,MHz, 170--200\,MHz, and 200--231\,MHz were cycled through sequentially, changing every two minutes. To cover the Dec range $-90^\circ$--$30^\circ$, observations were performed as a series of drift scans covering a single Declination per night. \Tab~\ref{tab:obs} summarises the vital statistics of these observations, which were all taken from the first year of GLEAM observations.

\begin{table}
\caption{GLEAM observations imaged in this paper. $N_\mathrm{flag}$ is the number of flagged tiles out of the 128~available. The calibrator is used to find initial bandpass and phase corrections as described in \Sect~\ref{sec:calibration}.\label{tab:obs}}
\begin{tabular}{ccccc}
\hline
Date & RA range (h) & Dec ($\arcdeg$) & $N_\mathrm{flag}$ & Calibrator\tabularnewline
\hline
\hline
2014-06-09 & 12--22 & $-27$ & 8 & 3C444\tabularnewline
2014-06-10 & 12--22 & $-40$ & 8 & 3C444\tabularnewline
2014-06-11 & 12--22 & $+2$ & 8 & Hercules A\tabularnewline
2014-06-12 & 12--18.5 & $-55$ & 8 & 3C444\tabularnewline
2014-06-13 & 12--19 & $-13$ & 8 & Centaurus A\tabularnewline
2014-06-14 & 12--22 & $-72$ & 9 & Hercules A\tabularnewline
2014-06-15 & 12--22 & $+18$ & 13 & Virgo A\tabularnewline
2014-06-16 & 18.5--22 & $-13$ & 8 & 3C444\tabularnewline
2014-06-18 & 18.5--22 & $-55$ & 8 & 3C444\tabularnewline
\hline
\end{tabular}
\end{table}

\subsection{Calibration}\label{sec:calibration}

Calibration is performed following the same method as \cite{2017MNRAS.464.1146H}: for each nightly drift scan, a bright calibrator source was observed (see \tab~\ref{tab:obs}) and the per-tile, per-polarisation, per-frequency channel amplitude and phase gains are calculated from that observation using \textsc{MitchCal} \citep{2016MNRAS.458.1057O}, using all baselines except the shortest ($<60$\,m). These gains are then applied to all observations in that drift scan.

For those observations where a bright source lies in the sidelobe of the primary beam, the visibilities are phase-rotated to the location of the source, and a peeling process performed. A model of the source is used to generate calibration solutions for that region of sky, and those gains applied to the model. This model is subtracted from the visibilities of the observation, which are then phase rotated back to the original pointing direction. In this way, the chromatic effect of the primary beam sidelobe is taken into account when removing the source, without distorting the overall gains of the observation.

At this stage, the processing diverges depending on whether the Galactic Plane lies within the main field-of-view of the observation. For observations of purely extragalactic sky, a self-calibration process is performed. \textsc{WSClean} is used to generate initial $XX$, $XY$, $YX$, and $YY$ images for each observation, over the full 30.72\,MHz, stopping the \textsc{Clean} process at the first negative component. These instrumental Stokes images are transformed into astronomical Stokes images by applying the MWA primary beam model by \cite{2017PASA...34...62S}. As the model sky should be unpolarised, $Q$, $U$, and $V$ were set to zero, while Stokes~$I$ is decomposed back into instrumental Stokes and used to predict a set of model visibilities, again using \textsc{WSClean}. \textsc{Mitchcal} is again used to generate a new set of calibration solutions (also without the shortest $<60$-m baselines), which are applied to each observation.

For those observations where the Galactic plane is within the field-of-view, the self-calibration process was not stable, and resulted in poorer-quality images, despite several attempts to find calibration and imaging settings to make this possible. Therefore instead, these images are calibrated using the most temporally adjacent self-calibration solution from the drift scan. Typically the largest amount of power in images of the Galactic plane is detected using the short baselines; these are minimally affected by the ionosphere due to their small separations. Low-frequency antenna gains also tend to vary only slowly as the instrument is very temperature-stable, and instrumental temperature is small compared to the sky. Therefore transferring calibration solutions usually results in good calibration fidelity except around the brightest and most compact sources.

\subsection{Imaging}

For those observations where the Galactic Plane was is a significant source of emission (i.e. appears at $<20$\,\% of the primary beam sensitivity), \textsc{WSClean} is used to generate images with the following settings:
\begin{itemize}
\item A SIN projection centred on the minimum-$w$ pointing, i.e. hour angle~$=0$, Dec~$-26.7^\circ$
\item four 7.68-MHz channels jointly-cleaned using the ``joinchannels'' option, which also produces a 30.72-MHz MFS image;
\item four instrumental Stokes images jointly-cleaned using the ``joinpolarisations'' option, in which peaks are detected in the summed combination of the polarisations, and components are refitted to each polarisation once a peak location is selected;
\item automatic thresholding down to $3\sigma$, where $\sigma$ is the rms of the residual MFS image at the end of each major cycle;
\item a major cycle gain of 0.85, i.e. 85\% of the flux density of the clean components are subtracted in each major cycle;
\item five or fewer major cycles, in order to prevent the occasional failure to converge during cleaning between 3 and 4$\sigma$;
\item $10^6$ minor cycles, a limit which is never reached;
\item $4000\times4000$ pixel images, which encompasses the field-of-view down to 10\% of the primary beam;
\item ``robust'' weighting of -1 \citep{1995AAS...18711202B}, which for this configuration of the MWA is a good trade-off between resolution and sensitivity; 
\item a frequency-dependent pixel scale such that each image always has 3.5--5 pixels per FWHM of the restoring beam;
\item a restoring beam of a 2-D Gaussian fit to the central part of the dirty beam, which is similar in shape (within 10\,\%) for each frequency band of the entire survey, but varies in size depending on the frequency of the observation.
\end{itemize}
To obtain the best images of the Galactic plane, the \textsc{WSClean} imaging strategy is also modified, by applying \texttt{multiscale} \textsc{Clean}, with the default deconvolution scale settings. We also lower the major cycle gain to 0.6, which reduces the number of detected clean components subtracted in each major cycle. This is done to eliminate a trap which multiscale clean sometimes experiences, where it subtracts components on one scale, only to add them back in again in a different scale, and begins to ``oscillate'' until the resulting images are no longer physical.

In order to be consistent with \cite{2017MNRAS.464.1146H}, the Molonglo Reference Catalogue at 408\,MHz \citep[MRC;][]{1981MNRAS.194..693L,1991Obs...111...72L} is then used to set a basic flux density scale for the snapshot images (assuming a spectral index $\alpha=-0.85$). We rescale the images by selecting a sample of sources and cross-matching them with MRC, then calculate the ratio between the measured flux densities and those predicted from MRC, and apply this ratio as a multiplier. Failing to do this would lead to flux density scale variations of order 10--20\% between snapshots.

\subsection{Astrometric calibration}\label{sec:astrometry}

The ionosphere introduces a $\lambda^2$-dependent position shift to the observed radio sources, which varies with position on the sky. Following the same method of \cite{2017MNRAS.464.1146H}, we use \textsc{fits\_warp} \citep{2018A&C....25...94H} to calculate a model of position shifts based on the difference in positions between the sources in the snapshot and those in a reference catalogue, and then use this model to de-distort the images. We use the same reference catalogue as \cite{2017MNRAS.464.1146H}: for Decs south of $18\fdg5$, MRC, and for those further north, a similar catalogue formed by cross-matching the NRAO VLA Sky Survey \citep[NVSS; ][]{1998AJ....115.1693C} at 1.4\,GHz, and the VLA Low-frequency Sky Survey Redux at 74\,MHz \citep[VLSSr; ][]{2014MNRAS.440..327L}. For this ancillary catalogue, we calculated a 408\,MHz flux density assuming a simple power law spectral index for every source ($S\propto\nu^\alpha$), and then discarding all sources with $S_\mathrm{408\,MHz}<0.67$\,Jy, the same minimum flux density as MRC. In each snapshot there are 100--1000 crossmatched sources, depending on observation quality and frequency, which results in high-quality de-distortion results, with residual astrometric differences of just 5--10" at the lowest frequencies, and 0.5--$2.5"$ at the highest frequencies.

\subsection{Mosaicking}\label{sec:mosaic}

Similarly to the original pipeline of \cite{2017MNRAS.464.1146H}, the polarisation calibration is not sufficient to make use of the cross-polarisation terms, and these are discarded at this stage. Basic checks for imaging quality are performed on the snapshots, and any with very high RMS are discarded ($\approx2$\%), leaving 11,802 $\times$ 7.68-MHz XX and YY snapshots covering the iG region. Following \cite{2017MNRAS.464.1146H}, we mosaic the XX and YY polarisations separately, normalise them to the same flux density scale using a Dec-dependent polynomial fit to the source flux densities, mosaic XX and YY together and applying a primary beam model to each, to make pseudo-Stokes-$I$. We then apply correction factors of order $\approx20$\,\% to fix residual uncertainties in the primary beam model, also derived by \cite{2017MNRAS.464.1146H}. This ensures that the iG region is on the same flux density scale as the oG region and the compact source catalogue.

For optimal signal-to-noise when mosaicking the night-long scans together, we use inverse-variance weighting, where the variance is calculated as the square of the RMS, calculated by \textsc{BANE}. However, \textsc{BANE}'s algorithm is optimised for images where more than half of the sampled area is noise-like. This assumption fails in the complex, confused Galactic Plane, causing the RMS to be strongly correlated with real structures, and thereby reducing the effectiveness of the inverse variance weighting in reducing the noise. To overcome this, we interpolate the RMS maps over $|b|<5^\circ$ using the \textsc{SciPy} function \textsc{interpolate.griddata} \citep{Jones_scipy_2001}.

After combining each drift scan and correcting their flux density scales, all nine scans are mosaicked together with inverse variance weighting to form one large mosaic for each 7.68-MHz frequency channel. At this stage, we also form a 60-MHz bandwidth ``wideband'' image over 170--231\,MHz, as this gives a good compromise between sensitivity and resolution, and will be used for source-finding (\Sect~\ref{sec:catalogue}. We also form three further 30.72-MHz images from 72--103\,MHz, 103--134\,MHz, and 139--170\,MHz, as the improved sensitivity of these images is useful for characterising SNR (Hurley-Walker et al. 2019b).

\subsection{Calculation of the PSF}\label{sec:psf}

Residual uncorrected ionospheric distortions can cause slight blurring of the final mosaicked point spread function (PSF). Again following \cite{2017MNRAS.464.1146H}, we can calculate the blurring effect by selecting unresolved sources via MRC and VLSSr, then calculate the corrected PSF by measuring the size and shape of these sources in the GLEAM mosaics. As with the RMS measurement made in \Sect~\ref{sec:mosaic}, this is unreliable over the iG region, so we interpolate the PSF maps over $|b|<5^\circ$ using the \textsc{SciPy} function \textsc{interpolate.griddata} \citep{Jones_scipy_2001}.

After the PSF map has been measured, its antecedent mosaic is multiplied by a (position-dependent) ``blur'' factor of
\begin{equation}
\centering
R = \frac{a_\mathrm{PSF} b_\mathrm{PSF}\cos\mathrm{ZA}}{a_\mathrm{rst} b_\mathrm{rst}}\label{eq:ionoblur}
\end{equation}
where $a_\mathrm{rst}$ and $b_\mathrm{rst}$ are the FWHM of the major and minor axes of the restoring beam, $a_\mathrm{PSF}$ and $b_\mathrm{PSF}$ are the FWHM of the major and minor axes of the PSF, and ZA is the zenith angle. This has the effect of normalising the flux density scale such that both peak and integrated flux densities agree, as long as the correct, position-dependent PSF is used \citep{2018PASA...35...11H}. Values of $R$ are typically 1.0--1.2.

\subsection{Final images}\label{sec:images}

\begin{sidewaysfigure*}
    \centering
    \includegraphics[width=\textwidth]{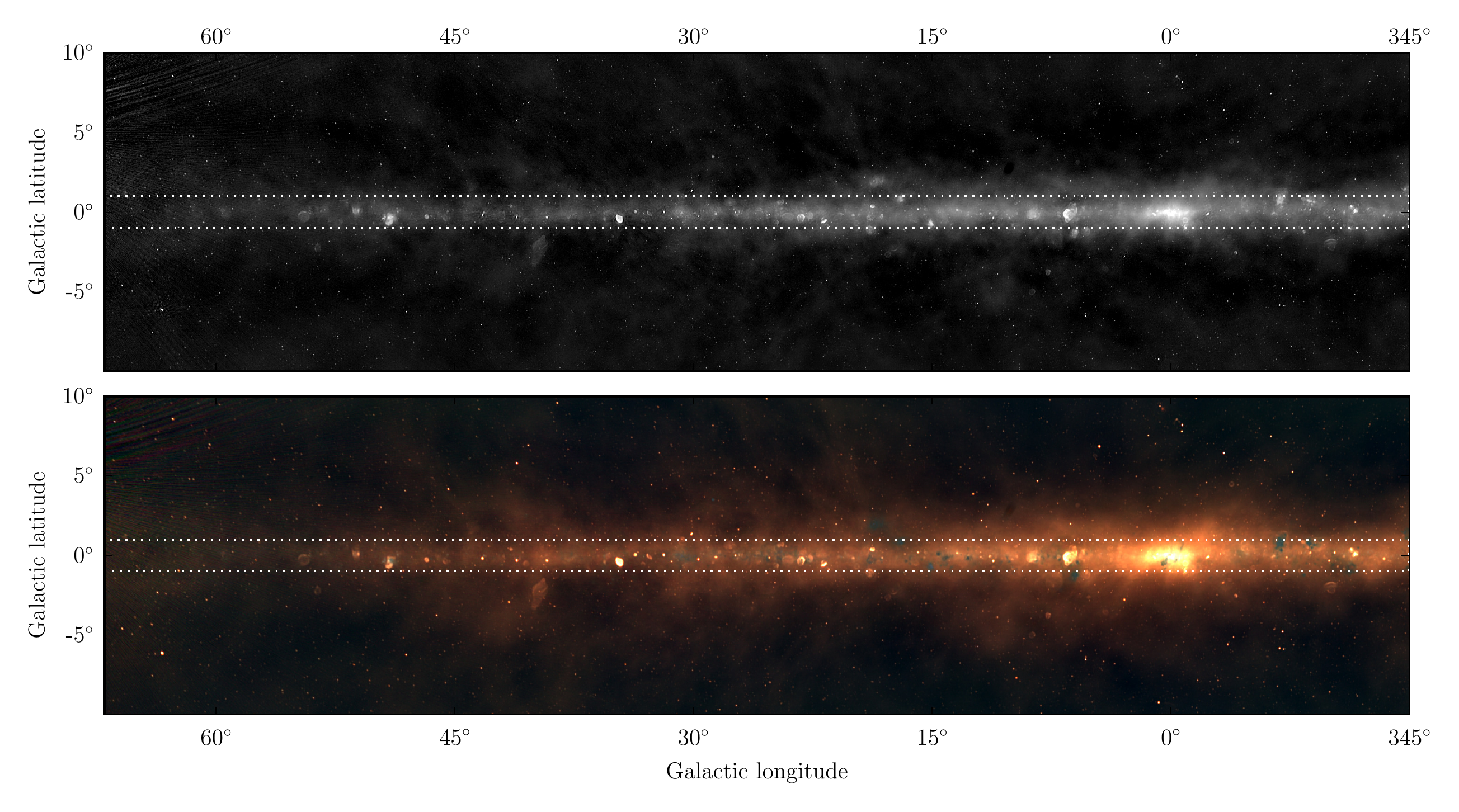} 
   \caption{The wide-bandwidth images from the data described in this paper; this figure shows the iG region.
    The top panel shows the 170--231\,MHz image which is used for source-finding (see \Sect~\ref{sec:catalogue}), between $-0.1$ and 5.0\,Jy\perbeam, with an $\arcsinh$ stretch.
    The bottom panel shows an RGB cube formed of the 72--103\,MHz (R), 103--134\,MHz (G), and 139--170\,MHz (B) data, between $-1$ and 10\,Jy\perbeam.
    Dotted white lines indicate $|b|=1^\circ$; source-finding is only performed outside of this region.
    }
    \label{fig:headline_week4}
\end{sidewaysfigure*}

\begin{sidewaysfigure*}
    \centering
    \includegraphics[width=\textwidth]{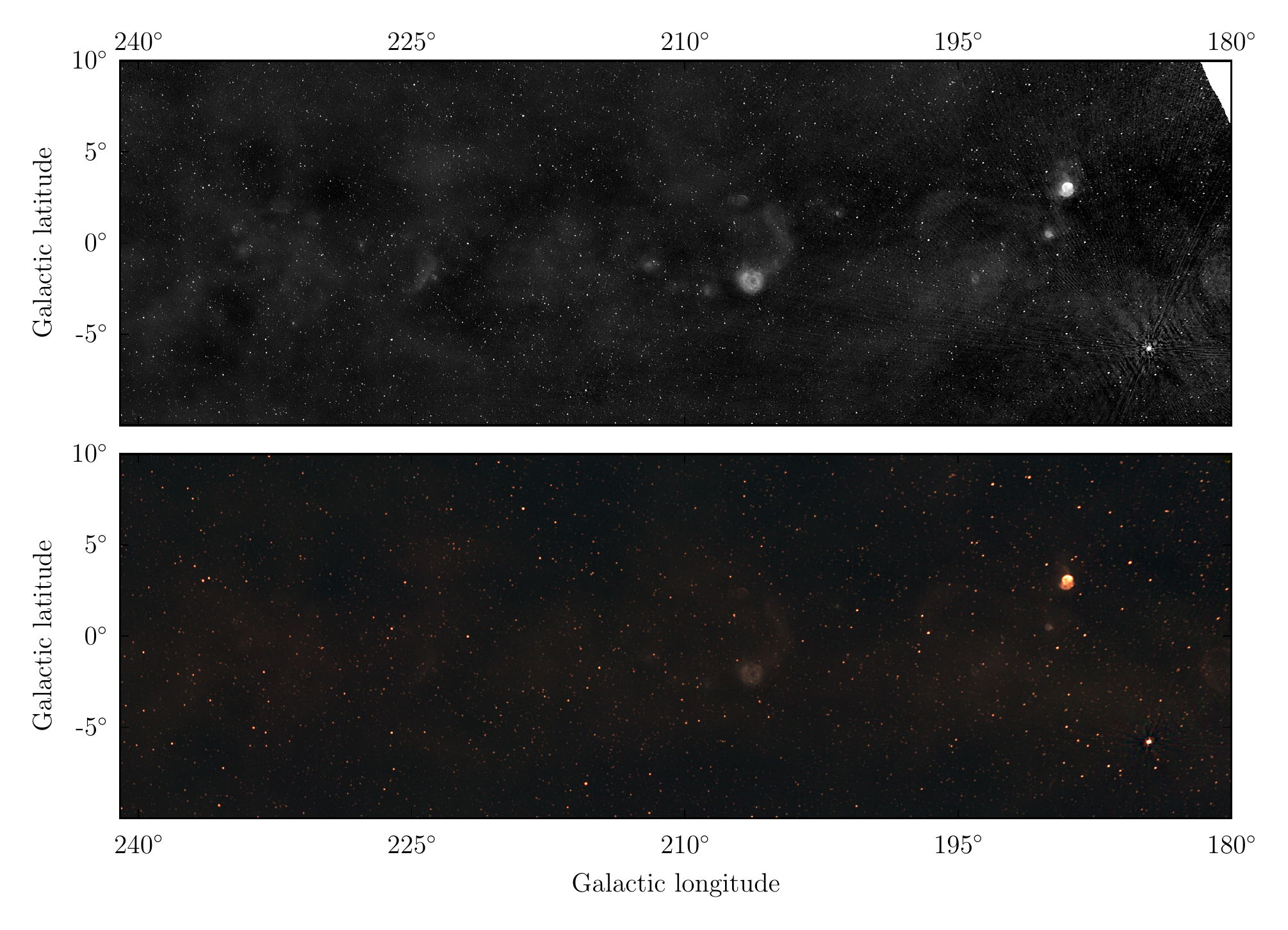} 
    \caption{
    The wide-bandwidth images from the data described in this paper; this figure shows the oG region.
    The top panel shows the 170--231\,MHz image which is used for source-finding (see \Sect~\ref{sec:catalogue}), between $-0.05$ and 1.5\,Jy\perbeam, with an $\arcsinh$ stretch.
    The bottom panel shows an RGB cube formed of the 72--103\,MHz (R), 103--134\,MHz (G), and 139--170\,MHz (B) data, between $-0.5$ and 5.0\,Jy\perbeam.
    }
    \label{fig:headline_week2}
\end{sidewaysfigure*}

The mosaicking stage of \sect~\ref{sec:mosaic} results in 21~mosaics, one with 60\,MHz bandwidth across 170--231\,MHz, and the other 20 covering the full bandwidth of 72--231MHz in 7.68-MHz narrow bands. Postage stamps of these images are available on both Skyview and the GLEAM website\footnote{http://mwatelescope.org/science/gleam-survey}. The header of every postage stamp contains the PSF information calculated in \Sect~\ref{sec:psf}, and the completeness information calculated in \Sect~\ref{sec:reliability}. \Figs~\ref{fig:headline_week4} and~\ref{fig:headline_week2} show the wide-bandwidth images from the data described in this paper.

\section{Compact source-finding}\label{sec:catalogue}

Source-finding is performed within $1\geq|b|\leq20^\circ$ for the iG, and $|b|\leq20^\circ$ for the oG. We avoid $|b|<1^\circ$ of the iG for several reasons: the presence of bright, large-scale, Galactic emission makes calculating accurate RMS and background images very difficult; the resolved emission is difficult to characterise as a set of elliptical Gaussians; and a catalogue of this region would be of limited further utility as fitted components would not often map directly to astrophysical objects, unlike compact sources, which tend to be radio galaxies, or compact Galactic objects such as pulsars, PWNe, (apparently) small SNR, and \textsc{Hii} regions.
 
Following the same strategy as \cite{2017MNRAS.464.1146H}: a deep wideband catalogue centred at 200\,MHz is formed, for sources with peak flux density $\geq5\times$ the local RMS. We use the ``priorised'' fitting technique to measure the flux densities of every detected source in the narrow-band images. The number of Gaussians allowed to form a source is limited to five, in order to prevent computationally expensive over-fitting of Galactic objects. We perform several checks on the quality of the catalogue, detailed below.

\subsection{Extended sources}

228~sources found during initial source-finding have a ratio of integrated to peak flux density greater than two. Visual inspection of these sources is performed and those which were components of large ($a\gtrsim10'$) objects are manually removed from the catalogue. These large objects are typically SNR (e.g. 3C\,400.2, Milne~56), and are not included in the catalogue as, like most of the objects in the region $|b|<1^\circ$ in the iG, they cannot be easily represented as a set of elliptical Gaussians. The number of sources manually removed from the catalogue in this way is \nremoved, leaving \nvresolved~sources with a ratio of integrated to peak flux density greater than two.

\subsection{Error derivation}

In this section we examine the errors reported in the catalogue. First, we examine the systematic flux density errors; then, we examine the noise properties of the wide-band source-finding image, as this must be close to Gaussian in order for sources to be accurately characterised, and for estimates of the reliability to be made, which we do in \Sect~\ref{sec:reliability}. Finally, we make an assessment of the catalogue's astrometric accuracy. These statistics are given in \Tab~\ref{tab:survey_stats}.

\begin{table*}
    \caption{Survey properties and statistics. We divide the survey into four parts, because the noise properties, and astrometric and flux calibration, differ slightly for each range. ``iG'' indicates the inner Galactic; ``oG'' the outer Galactic region; ``S'' indicates $-72\arcdeg\leq\mathrm{Dec}<+18\fdg5$; ``N'' indicates $\mathrm{Dec}\geq18\fdg5$. Values are given as the mean$\pm$the standard deviation. The statistics shown are derived from the wideband (200\,MHz) image. The flux density scale error applies to all frequencies, and shows the degree to which GLEAM agrees with other published surveys. The internal flux density scale error also applies to all frequencies, and shows the internal consistency of the flux density scale within GLEAM.\label{tab:survey_stats}}
    \begin{tabular}{ccccc}
    \hline
    Property & iG S & oG S & iG N & oG N \\
    \hline
    Number of sources & 7,722 & 12,534 & 665 & 1,176\\
    RA astrometric offset (") & \multicolumn{2}{c}{$-0\farcs4\pm3\farcs1$} &  \multicolumn{2}{c}{$-0\farcs1\pm1\farcs6$} \\
    Dec astrometric offset (") & \multicolumn{2}{c}{$-1\farcs1\pm3\farcs5$} &  \multicolumn{2}{c}{$1\farcs9\pm3\farcs0$} \\
    External flux density scale error (\%) & \multicolumn{2}{c}{8} & \multicolumn{2}{c}{13} \\
    Internal flux density scale error (\%) & \multicolumn{2}{c}{2} & \multicolumn{2}{c}{3} \\
    Completeness at 50\,mJy (\%) & 2 & 45 & 0 & 1 \\
    Completeness at 100\,mJy (\%) & 31 & 84 & 3 & 21 \\
    Completeness at 160\,mJy (\%) & 81 & 92 & 25 & 60 \\
    Completeness at 0.5\,Jy (\%) & 99 & 99 & 97 & 98 \\
    Completeness at 1\,Jy (\%) & 100 & 100 & 100 & 99 \\
    RMS (mJy\,beam$^{-1}$) & $22\pm22$ & $14\pm20$ & $10\pm5$ & $27\pm35$ \\
    PSF major axis (") & \multicolumn{2}{c}{$140\pm10$} & \multicolumn{2}{c}{$192\pm14$} \\
    PSF minor axis (") & \multicolumn{2}{c}{$131\pm4$} & \multicolumn{2}{c}{$135\pm2$} \\
    \hline
    \end{tabular}
\end{table*}

\subsubsection{Flux densities}
The errors on the flux density measurements arise from the Dec-dependent flux density scale correction derived by \cite{2017MNRAS.464.1146H}, as well as fitting errors as estimated by \textsc{Aegean}. As the scale correction is identical to that performed by \citeauthor{2017MNRAS.464.1146H}, we adopt the same uncertainty values of 8\,\% for $\mathrm{Dec} < +18.5^\circ$ and 13\,\% for $\mathrm{Dec} > +18.5^\circ$. These values are listed in the catalogue for each source, and should be added in quadrature with the fitting errors when comparing with other catalogues. When fitting solely within the GLEAM catalogue, the internal flux density scale errors of  2\,\% for $\mathrm{Dec} < +18.5^\circ$ and 3\,\% for $\mathrm{Dec} > +18.5^\circ$ should be used \citep[see ][for a more complete discussion]{2017MNRAS.464.1146H}.

\subsubsection{Astrometry}\label{sec:overall_astrometry}

Following \cite{2017MNRAS.464.1146H}, we measure the astrometry using the 200-MHz catalogue, as this provides the locations and morphologies of all sources. To determine the astrometry,
unresolved ($(a\times b)/(a_\mathrm{PSF}\times b_\mathrm{PSF})<1.1$),
isolated (no internal match within 10') GLEAM sources are cross-matched with similarly isolated sources in the NVSS and the Sydney University Molonglo Sky Survey \citep[SUMSS;][]{1999AJ....117.1578B}; the positions of sources in these catalogues are assumed to be correct and RA and Dec offsets are measured with respect to those positions. For $\mathrm{Dec}\leqslant+18\fdg5$, the average RA offset is $-0\farcs4\pm3\farcs1$, and the average Dec offset is $-1\farcs1\pm3\farcs5$. North of $+18\fdg5$, the average RA offset is $-0\farcs1\pm2\farcs7$ and the average Dec offset is $1\farcs9\pm3\farcs0$. These offsets may be somewhat different because a modified VLSSr/NVSS catalogue was used to replace MRC North of its Dec limit of 18\fdg5 for astrometric calibration (see \sect~\ref{sec:astrometry}). These offsets are all completely consistent with the values published by \cite{2017MNRAS.464.1146H} for the majority of the extragalactic sky.

\begin{figure}
    \centering
   \includegraphics[width=0.5\textwidth]{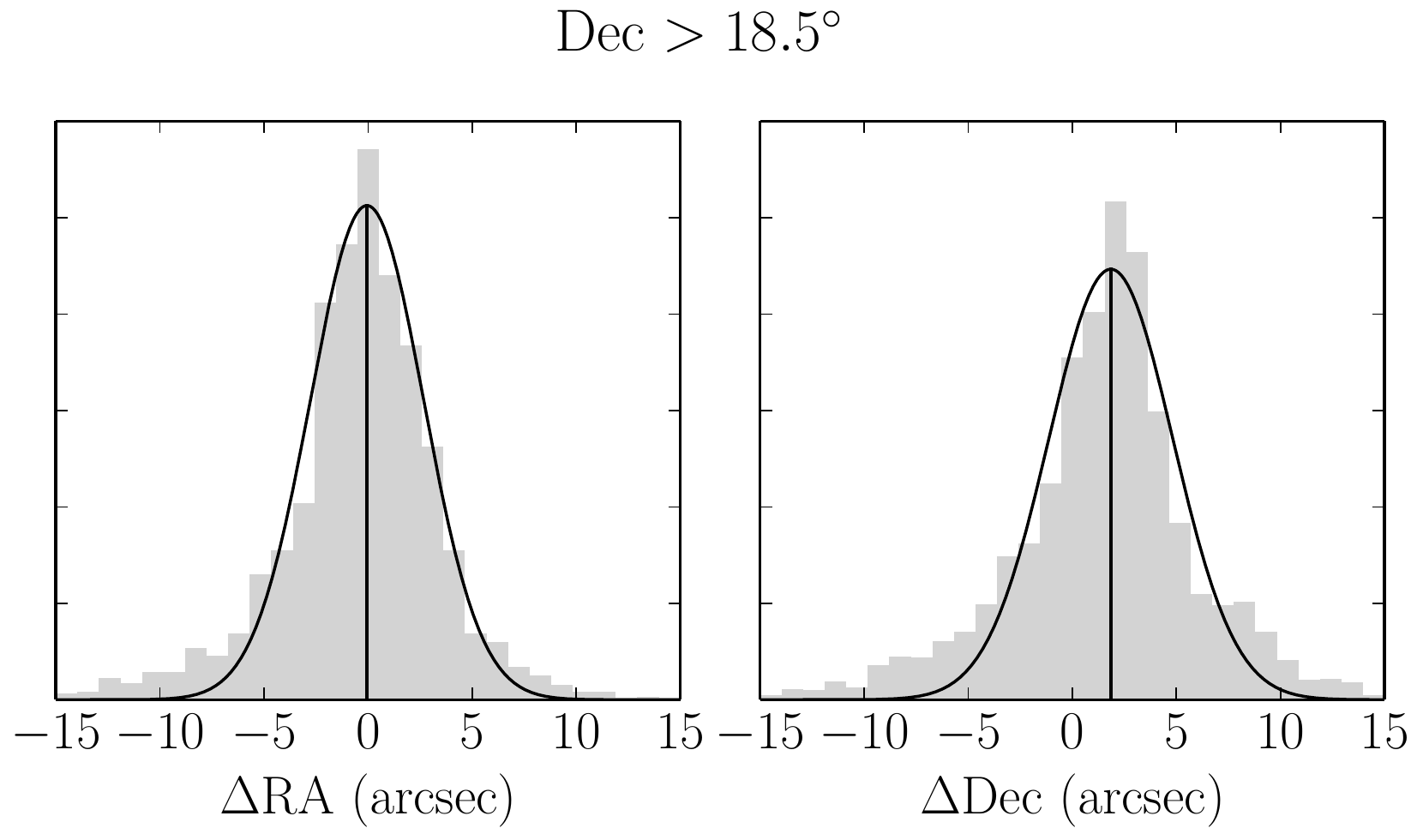}
   \includegraphics[width=0.5\textwidth]{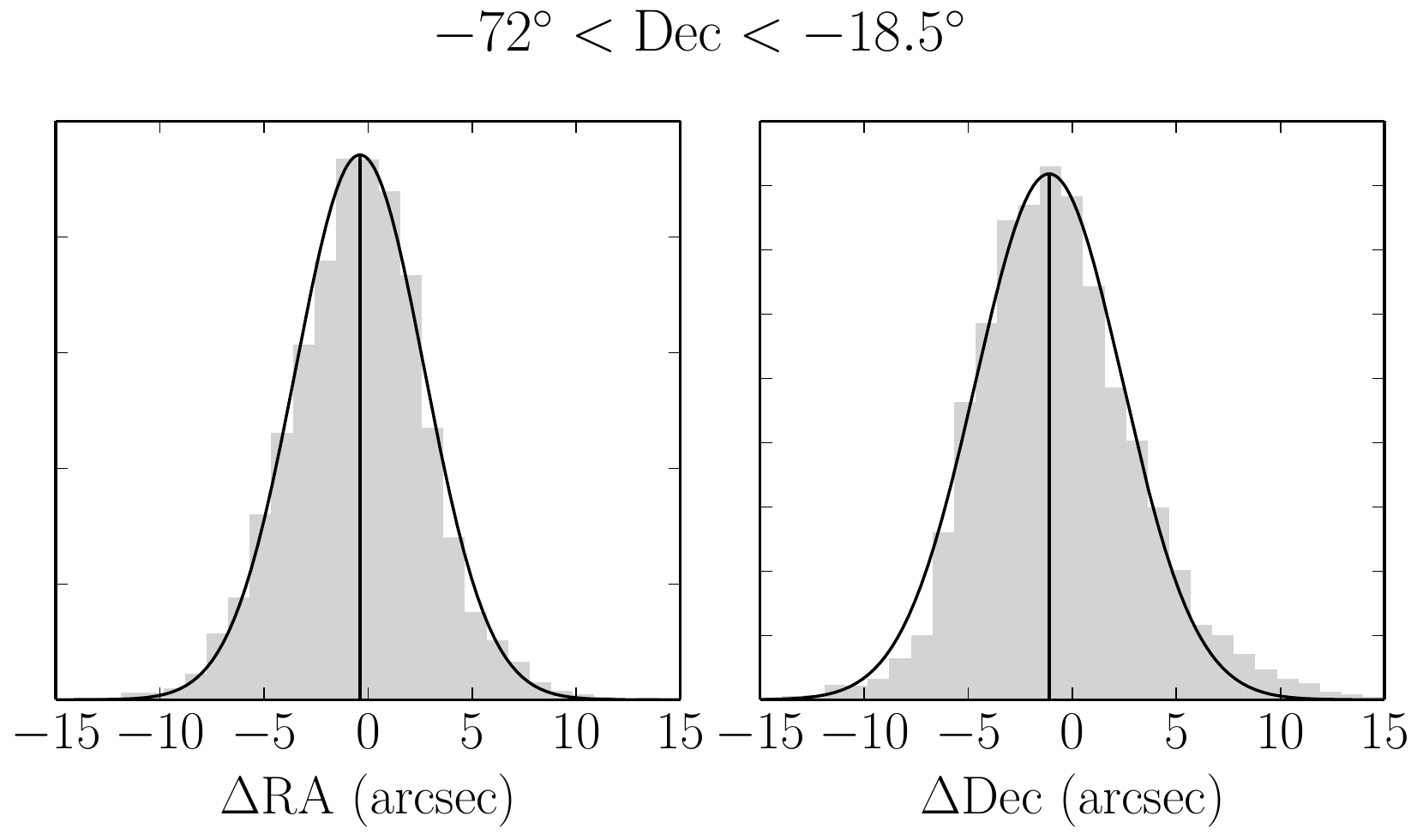}
    \caption{Histograms, weighted by source S/N, of astrometric offsets, for isolated compact GLEAM sources crossmatched with NVSS and SUMSS as described in \sect~\ref{sec:overall_astrometry}. The black curves show Gaussian fits to each histogram. Solid vertical lines indicate the mean offsets. The top panel shows sources on the northern edge of the survey, $\mathrm{Dec}\geq+18\fdg5$ and the bottom panel shows sources south of this cut-off.}
    \label{fig:overall_astrometry}
\end{figure}

In 99\% of cases, fitting errors are larger than the measured average astrometric offsets. Given the scatter in the measurements, we do not attempt to make a correction for these offsets. As each snapshot has been corrected, residual errors should not vary on scales smaller than the size of the primary beam. \fig~\ref{fig:overall_astrometry} shows the density distribution of the astrometric offsets, and Gaussian fits to the RA and Dec offsets, which were used to calculate the values listed in this section.

\subsubsection{Noise properties}\label{sec:noise_properties}

\begin{figure}
    \centering
   \includegraphics[width=0.5\textwidth]{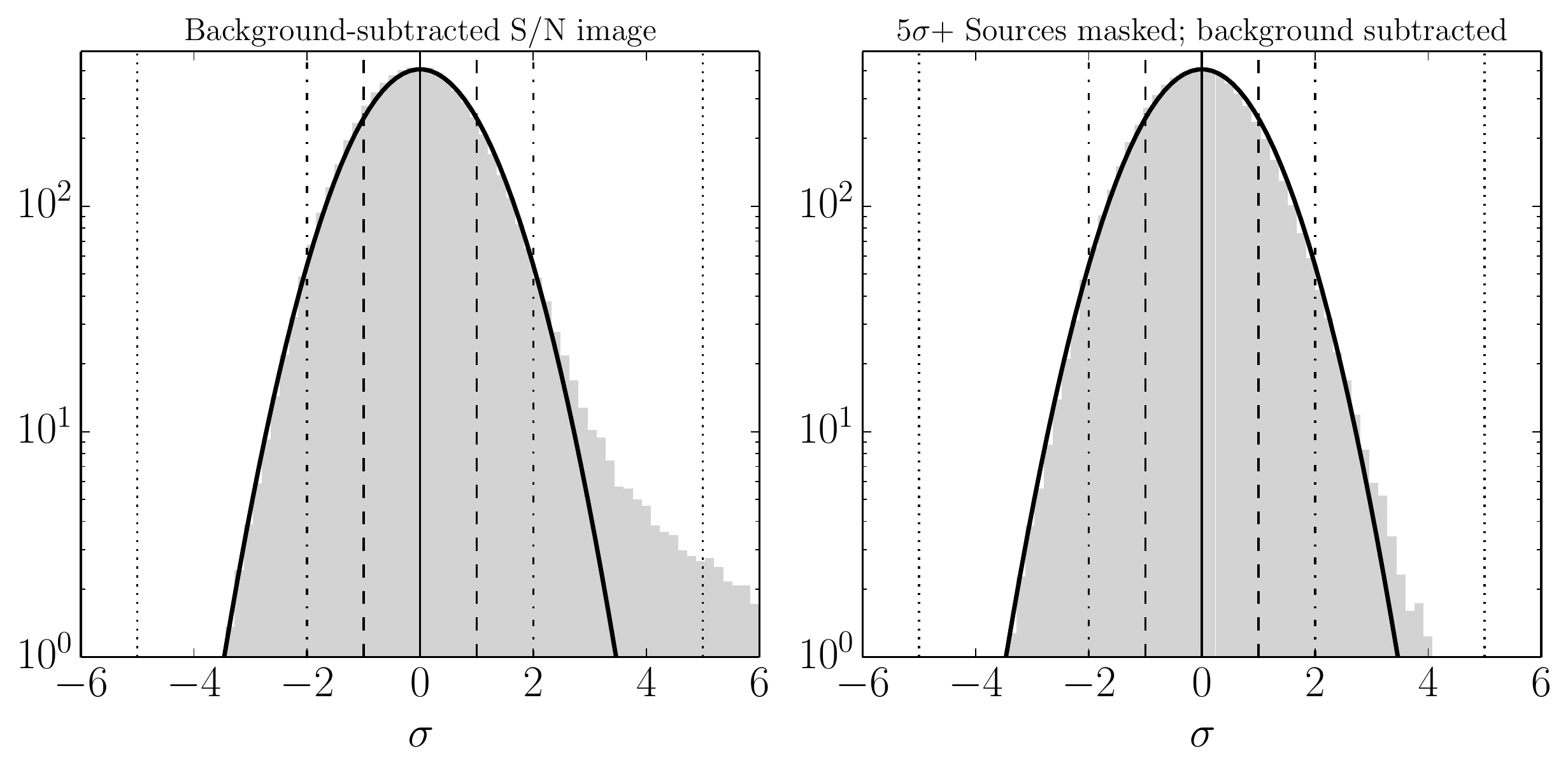}
    \caption{Noise distribution in a typical 25\,square degrees of the wideband source-finding image. \textsc{BANE} measures the average RMS in this region to be 24\,mJy\,beam$^{-1}$. To show the deviation from Gaussianity, the ordinate is plotted on a log scale. The leftmost panel shows the distribution of the S/Ns of the pixels in the image produced by subtracting the background and dividing by the RMS map measured by \textsc{BANE}; the right panel shows the S/N distribution after masking all sources detected at $5\sigma$ down to $0.2\sigma$. The light grey histograms show the data. The black lines show Gaussians with $\sigma=1$; vertical solid lines indicate the mean values. $|\mathrm{S/N}|=1\sigma$ is shown with dashed lines, $|\mathrm{S/N}|=2\sigma$ is shown with dash-dotted lines, and $|\mathrm{S/N}|=5\sigma$ is shown with dotted lines.}
    \label{fig:noise_distribution}
\end{figure}

We briefly examine the noise properties of the wideband (200-MHz) image. We use a 25\,deg$^2$ region covering $-10^\circ<b<-1^\circ$ with fairly typical source distribution and a background which slowly varies on $\approx5^\circ$ scales due to the undeconvolved large-scale sidelobes of the Galactic Plane. Following \cite{2017MNRAS.464.1146H}, we measure the background of the region using \textsc{BANE}, and subtract it from the image. We then use \textsc{AeRes} from the \textsc{Aegean} package to mask out all sources which were detected by \textsc{Aegean}, down to $0.2\times$ the local RMS. Histograms of the remaining pixels are shown, for the unmasked and masked images, in \Fig~\ref{fig:noise_distribution}.

The noise level in the selected region is $\approx3\times$ higher than the RMS of the wideband image produced for the extragalactic sky \citep{2017MNRAS.464.1146H}. This is mainly due to the increased overall system temperature due to the high sky temperature. Confusion therefore forms a smaller fraction of the noise contribution, and thus the noise distribution is almost completely symmetric. For regions with lower noise, the distribution will start to skew positive, as seen by \citeauthor{2017MNRAS.464.1146H}. Noise and background maps are made available as part of the survey data release.

\subsection{Completeness and reliability}\label{sec:reliability}

Following the same procedure as \cite{2017MNRAS.464.1146H}, simulations are used to quantify the completeness of the source catalogue at 200\,MHz, using the wideband mosaics. 
Thirty-three realisations are used in which 28,000 simulated point sources of the same flux density were injected into the 170--231\,MHz mosaics, between $1^\circ\leq|b|\leq10^\circ$.
The flux density of the simulated sources is different for each realisation, spanning the range 25~mJy to 1~Jy.
The positions of the simulated sources are chosen randomly but not altered between realisations; to avoid introducing an artificial factor of confusion in the simulations, simulated sources are not permitted to lie within $10'$ of each other\footnote{This would lead to an overestimate of the completeness for regions where the source density is greater than 36 sources per square degree, but even in the deepest extragalactic fields it reaches only 20 sources per square degree, and in the iG and oG regions it is more typically 10 sources per square degree.}. Sources are injected into the mosaics using \textsc{AeRes}. The major and minor axes of the simulated sources are set to $a_\mathrm{psf}$ and $b_\mathrm{psf}$, respectively.

For each realisation, the source-finding procedures described in \Sect~\ref{sec:catalogue} are applied to the mosaics and the fraction of simulated sources recovered is calculated.
In cases where a simulated source is found to lie too close to a real ($>5\sigma$) source to be detected separately, the simulated source is considered to be detected if the recovered source position is closer to the simulated rather than the real source position. This type of completeness simulation therefore accounts for sources that are omitted from the source-finding process through being too close to a brighter source. In crowded regions with medium-scale features which may not be detected as discrete sources, the simulated sources are less likely to be detected, so the completeness may be underestimated in these areas, e.g. $1^\circ<|b|<2^\circ$ of the iG.

\begin{figure}
\begin{center}
\includegraphics[width=0.35\textwidth, angle=270]{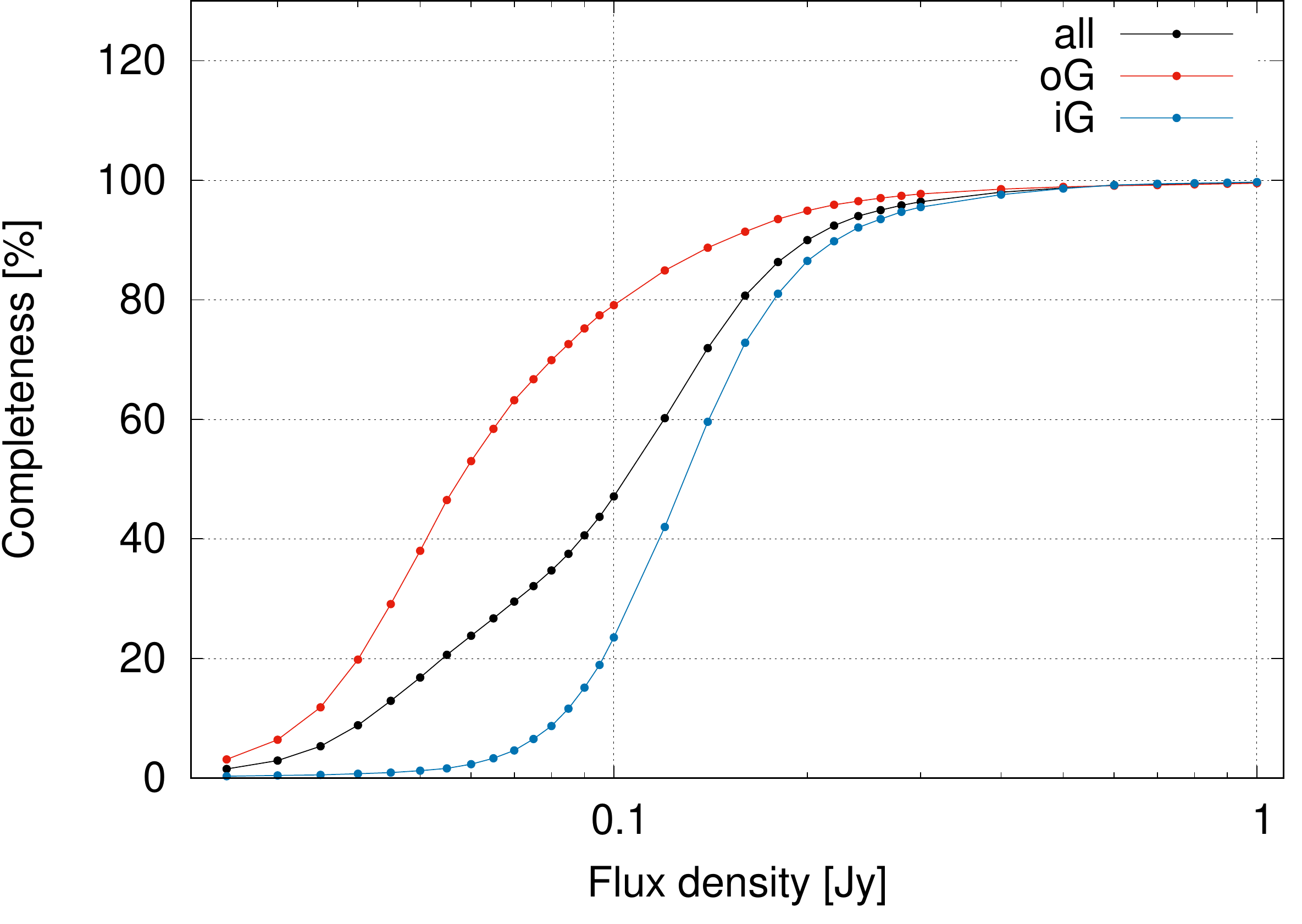} 
\caption {Estimated completeness of the catalogue for $1\fdg5<|b|<10^\circ$ as a function of $S_{200 \mathrm{MHz}}$ in the iG (blue circles), in the oG (red circles), and overall (black circles).}
\label{fig:cmp_vs_s}
\end{center}
\end{figure}

\fig~\ref{fig:cmp_vs_s} shows the fraction of simulated sources recovered as a function of
$S_{200 \mathrm{MHz}}$ in the iG and oG regions of the Galactic Plane.
For the oG, the completeness is estimated to be 50\,\% at $\approx 60$\,mJy rising to
90\,\% at $\approx 200$\,mJy; these statistics are similar to the extragalactic catalogue \citep{2017MNRAS.464.1146H}. For the iG, the completeness is considerably worse: 50\,\% at $\approx 120$\,mJy rising to 90\,\% at $\approx 220$\,mJy. Errors on the completeness estimate are derived assuming Poisson errors on the number of simulated sources detected.

The completeness at any pixel position is given by $C = N_{\mathrm{d}}/N_{\mathrm{s}}$, where $N_{\mathrm{s}}$ is the number of simulated sources in a circle of radius $6\arcdeg$ centred on the pixel and $N_{\mathrm{d}}$ is the number of simulated sources that were detected above $5\sigma$ within this same region of sky.
The completeness maps, in \textsc{fits} format, can be obtained from the supplementary material. Postage stamp images from the GLEAM VO server also include this completeness information in their headers.

Again using the same procedure as \cite{2017MNRAS.464.1146H}, we use the same source-finding algorithm but invert the brightness, looking only for sources with $S_\mathrm{200MHz}<-5\sigma$. In the \catarea~deg$^2$ region, we find \nneg~negative sources. As the noise distribution is close to symmetric (\Sect~\ref{sec:noise_properties}), we expect to see an approximately equal number of false positive sources in the same area. We thus estimate the catalogue reliability to be:

$1.0 - \frac{\nneg}{\nsrc} = {\pctreliable}$\,\%

\subsection{Spectral fitting}\label{sec:alpha}

Following \cite{2017MNRAS.464.1146H}, we fit spectral energy distributions (following $S\propto\nu^\alpha$) to the twenty narrow-band flux density measurements for all detected sources. We retained fit parameters only for those sources with reduced~$\chi^2<1.93$, indicating a likelihood of correct fit $>99$\,\% for 18~degrees of freedom.

\begin{figure}
    \centering
   \includegraphics[width=0.5\textwidth]{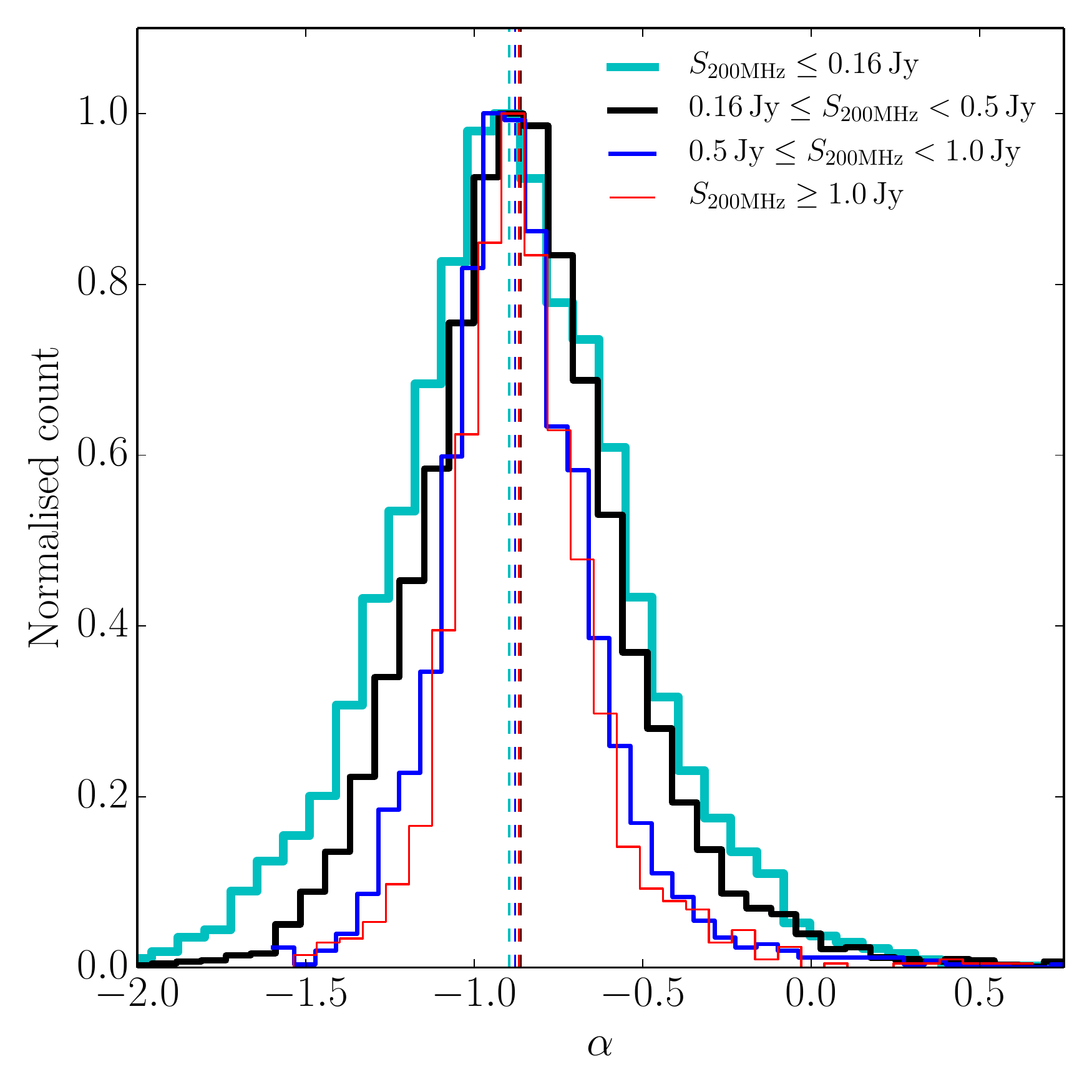}
    \caption{The spectral index distribution calculated for sources $|b|<10^\circ$, where the fit was successful (reduced~$\chi^2<1.93$). The cyan line shows sources with $S_\mathrm{200MHz}<0.16$\,Jy, the
black line shows sources with $0.16\leq S_\mathrm{200MHz}<0.5$\,Jy, the blue line shows sources with $0.5\leq S_\mathrm{200MHz}<1.0$\,Jy, and the red line shows sources with $S_\mathrm{200MHz}>1.0$\,Jy. The dashed vertical lines of the same colours show the median values for each flux density cut: $-0.89$, $-0.86$, $-0.88$, and $-0.87$, respectively.
    \label{fig:alpha_distribution}}
\end{figure}

Despite the reduced number of measurements ($10\times$~fewer than \cite{2017MNRAS.464.1146H}), we are able to recover similar distributions of spectral indices, plotted in \Fig~\ref{fig:alpha_distribution}. The median fitted spectral indices are slightly steeper than those of \citeauthor{2017MNRAS.464.1146H}, even for identical flux density bins; e.g. for (high S/N) sources with $S_\mathrm{200MHz}>1$\,Jy, the median $\alpha$ has steepened from $-0.83$ to $-0.89$.

A potential explanation for this steepening is that pulsars, which have steep spectra of $\alpha=-1.8\pm0.2$, make up an increasingly large fraction of our catalogue at low Galactic latitudes. \cite{2018MNRAS.474.5008D} combined the NVSS at 1.4GHz with the Alternative Data Release of the Tata Institute for Fundamental Research GMRT Sky Survey \citep[TGSS-ADR1; ][]{2017A&A...598A..78I} at 150\,MHz to form a spectral index catalogue covering 80\,\% of the sky. They found an excess of 86 compact and 49 non-compact steep-spectrum sources in $|b|<10^\circ$ beyond what would be expected extrapolating from the extragalactic sky.

If we restrict our search to the brightest sources where $S_\mathrm{200MHz}>1.0$\,Jy, of which there are 1,399 within $|b|<10^\circ$, we would need 56 to be pulsars (with $\alpha=-1.8$) to shift the median $\alpha$ from $-0.83$ to $-0.89$. However, there is only a single source that meets these criteria in our catalogue, so it appears an excess population of very steep-spectrum sources does not explain the shift in $\alpha$. Since the significance of the shift of $\alpha$ is not large compared to the IQR of $0.22$, we are unable to determine from these data alone what the cause of the steepening may be. Examining the nature of individual sources by comparison with other data at other frequencies will likely reveal whether the shift is astrophysical or some as-yet uncharacterised systematic.

\subsubsection{Comparison with extragalactic catalogue}

\begin{figure*}
    \centering
   \includegraphics[width=\textwidth]{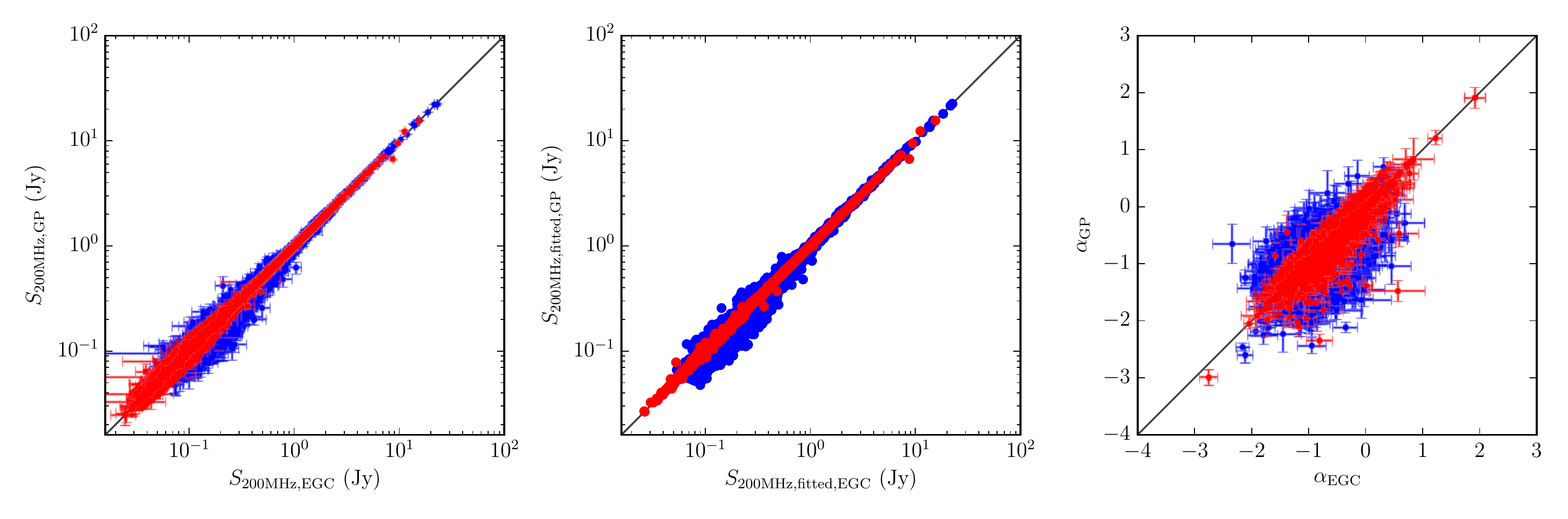}
    \caption{Three comparisons of the Galactic Plane catalogue (ordinates) with the extragalactic catalogue (abscissae). The left panel shows the integrated flux densities measured at 200\,MHz in the wideband mosaics, with error bars indicating only the fitting errors produced by \textsc{Aegean}. The middle panel shows the fitted 200-MHz flux densities over all twenty flux density measurements (see \Sect~\ref{sec:alpha}); error bars are not shown as they become very large on a log scale at low flux densities. The right panel shows the fitted $\alpha$. Red points are from the oG region (in which no additional data reduction was performed) and blue points indicate the iG region (which was completely reprocessed using more observations and multiscale \textsc{clean}). }
    \label{fig:comparison_egc}
\end{figure*}

Having reprocessed all data within $|b|<20^\circ$, there exists an overlap with the extragalactic catalogue of \cite{2017MNRAS.464.1146H}, for the range $10\arcdeg<|b|<20\arcdeg$. We can use this overlap to check for flux density scale consistency between the two catalogues. \Fig~\ref{fig:comparison_egc} shows a comparison of three major attributes between the catalogues: the integrated flux density in the 170--231\,MHz wideband images, the fitted 200\,MHz flux density over all frequency bands, and the fitted spectral index $\alpha$ over all frequency bands.

No biases or trends are visible; the catalogues are on the same flux density scales. There is a larger amount of scatter on the iG points, likely due to the increased number of observations used to generate the mosaics, and the slightly different processing scheme. As expected, there is very little scatter on the oG points, since the results of this source-finding are essentially the same as the measurements made by \cite{2017MNRAS.464.1146H}. Differences mainly arise in the range $b<-10^\circ$ for the oG, where a different set of mosaics was originally used by \cite{2017MNRAS.464.1146H} to generate the catalogue for that area of sky.

\subsection{Final catalogue}

Having established the quality of the catalogue, we filter it to retain only $|b|\leq10^\circ$, in order not to duplicate existing results. The resulting catalogue consists of \nsrc~radio sources detected over \catarea~deg$^2$. Of these, \nresolved~sources are resolved (ratio of integrated to peak flux density $>1.1$); just \nvresolved~sources are appreciably extended (ratio of integrated to peak flux density $>2.0$). \nfit~sources are fit well by power-law SEDs. The catalogue has \ncol columns identical to those in Appendix~A of \cite{2017MNRAS.464.1146H} and is available via Vizier. 

\begin{figure*}
    \centering
    \includegraphics[width=0.5\textwidth]{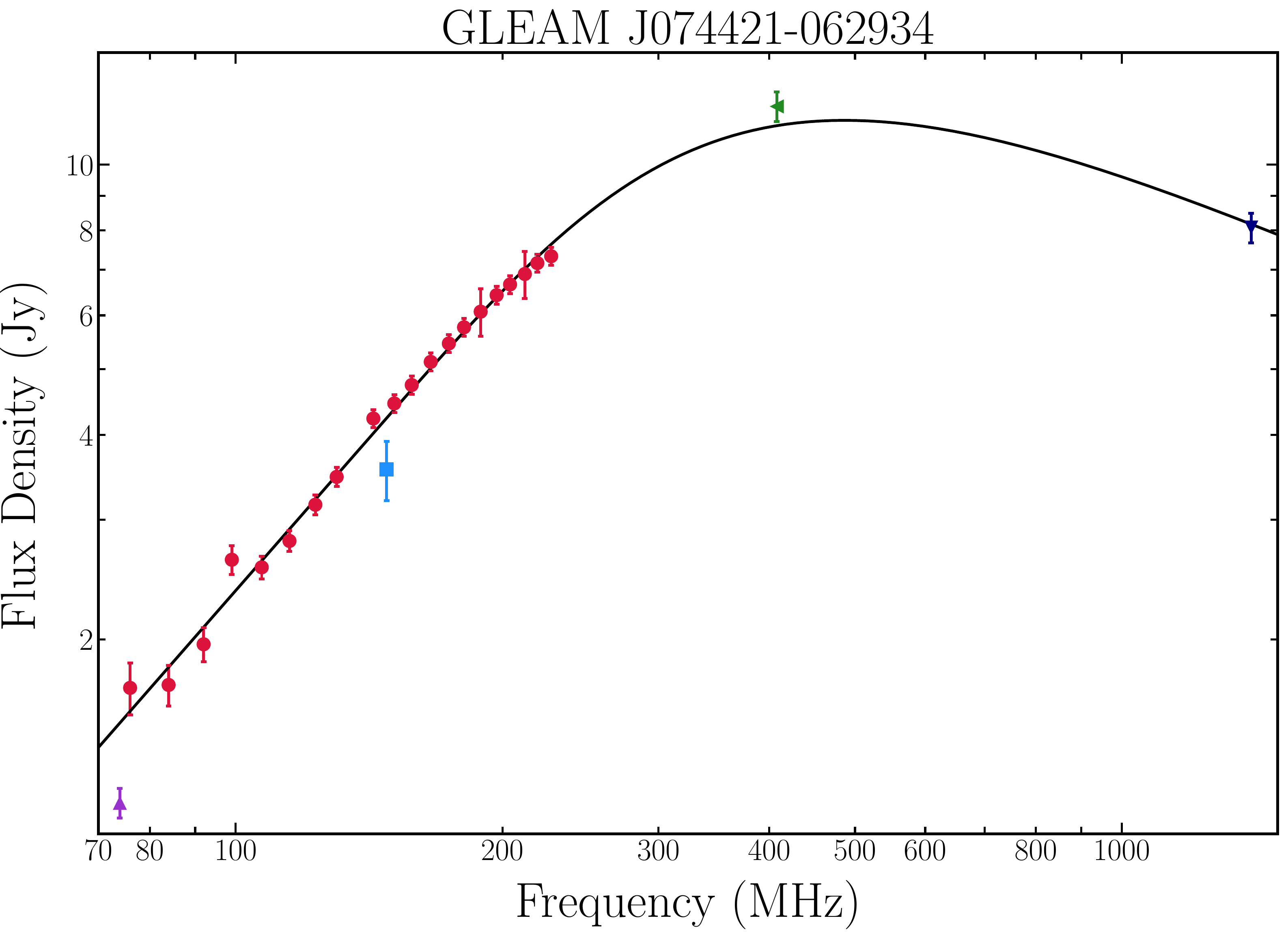}\includegraphics[width=0.5\textwidth]{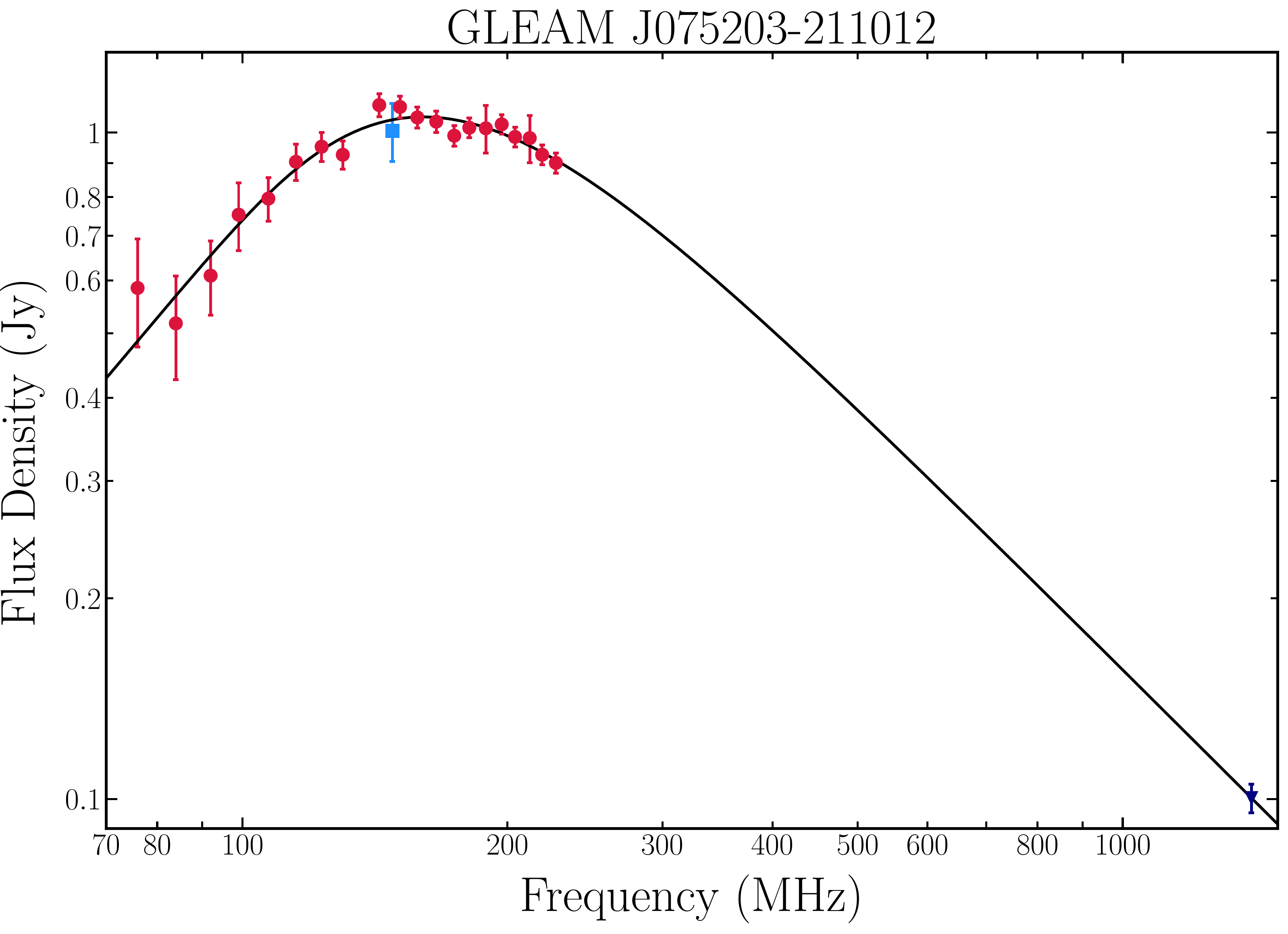} 
    \includegraphics[width=0.5\textwidth]{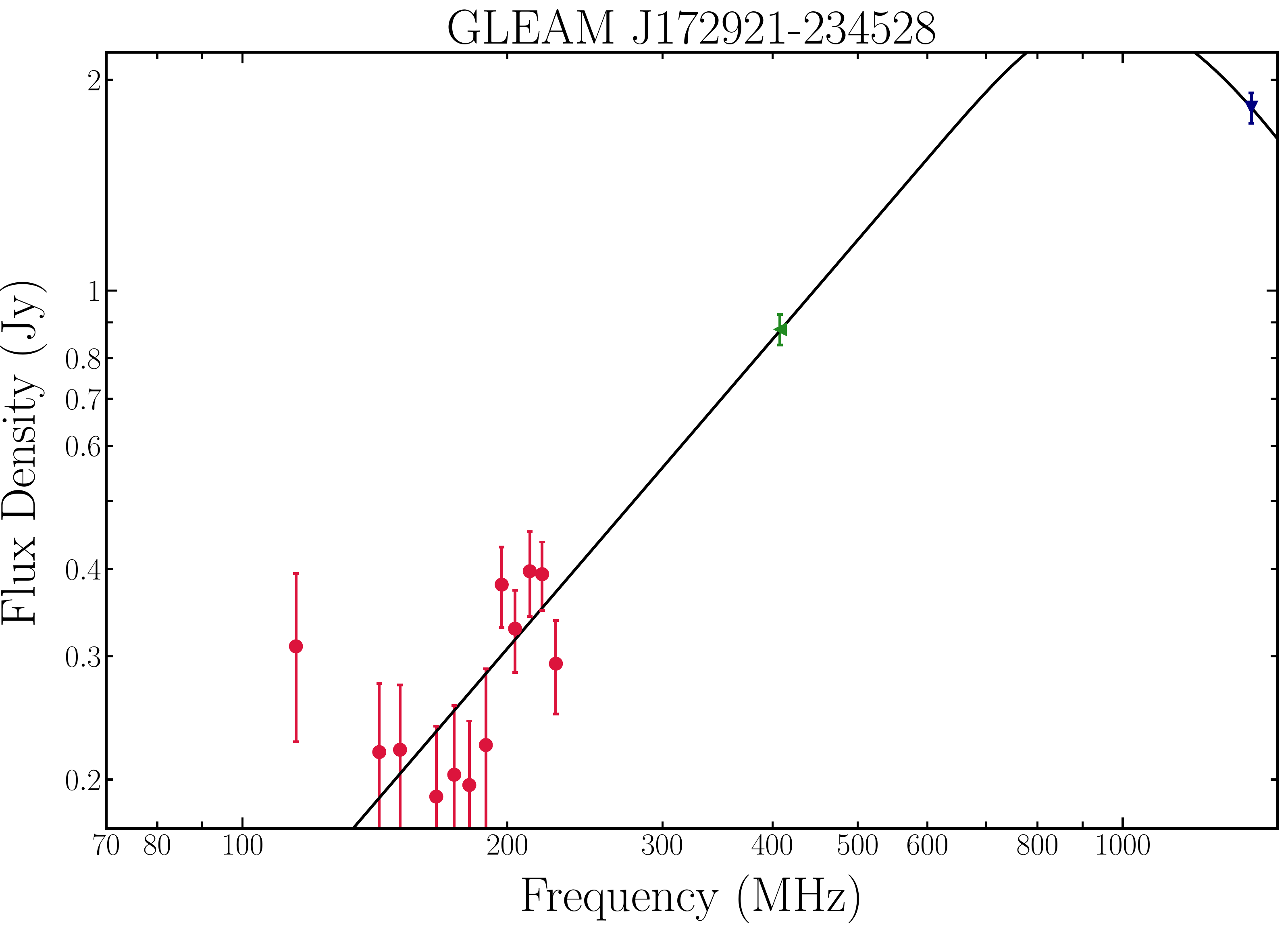}\includegraphics[width=0.5\textwidth]{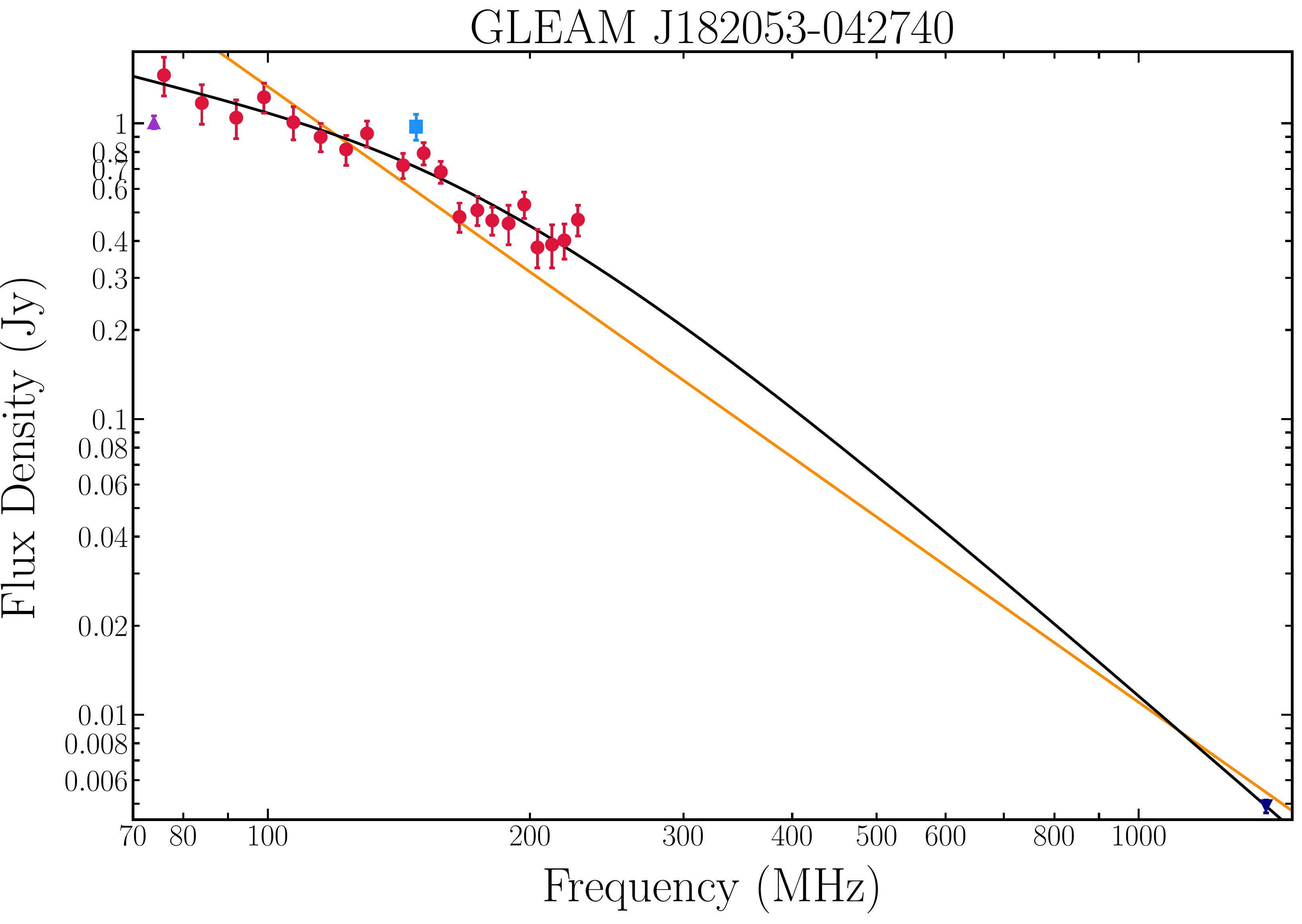}  
    \caption{Four example curved spectral energy distributions fit (black lines) to the narrow-band GLEAM measurements (red circles) and, where available, VLSSr (purple upward-pointing triangle), TGSS-ADR1 (light blue square), MRC (green leftward-pointing triangle), and NVSS (dark blue downward-pointing triangle), with fitting performed as by \protect\cite{Callingham2015}. The yellow line in the bottom-right panel shows a simple power-law SED fit. Sources shown are, from top-left to bottom-right, the known peaked-spectrum source 4C\,$-06.18$ \protect\citep{1967BAN....19..201D}, a previously-unknown peaked-spectrum source, the planetary nebula NGC~6369 \protect\citep{1918PLicO..13...55C}, and the pulsar J1820-0427 \protect\citep{1969Natur.222..963V}.  }
    \label{fig:seds}
\end{figure*}

The catalogue measurements can be used to perform more complex spectral fits, especially in conjunction with other radio measurements. \Fig~\ref{fig:seds} shows four example curved fits across GLEAM and data from the literature for two peaked-spectrum sources, a planetary nebula, and a pulsar.

\section{Galactic plane}\label{sec:galactic}

Utilising the large number of short baselines in the Phase~I configuration of the MWA, GLEAM has sensitivity to structures on large scales. The shortest baseline in the array is 7.72\,m in length, allowing access to angular scales of $<29^\circ$ at 76\,MHz, or $<10^\circ$ at 227\,MHz. Objects smaller than this should have correct flux densities across all frequencies. However, an important quantity which also changes with frequency is the background to any object to be measured, and this will also change rapidly with frequency due to the intrinsic synchrotron spectrum of the diffuse background ($T\propto\nu^{-2.7}$) and the aforementioned resolution effects. The point spread function will also change as a function of $\nu$, and is provided in the header in any postage stamp downloaded from the GLEAM VO server. We thus urge the reader to be careful of these effects when measuring flux densities in the images.

\begin{figure*}
    \centering
      \includegraphics[width=\textwidth]{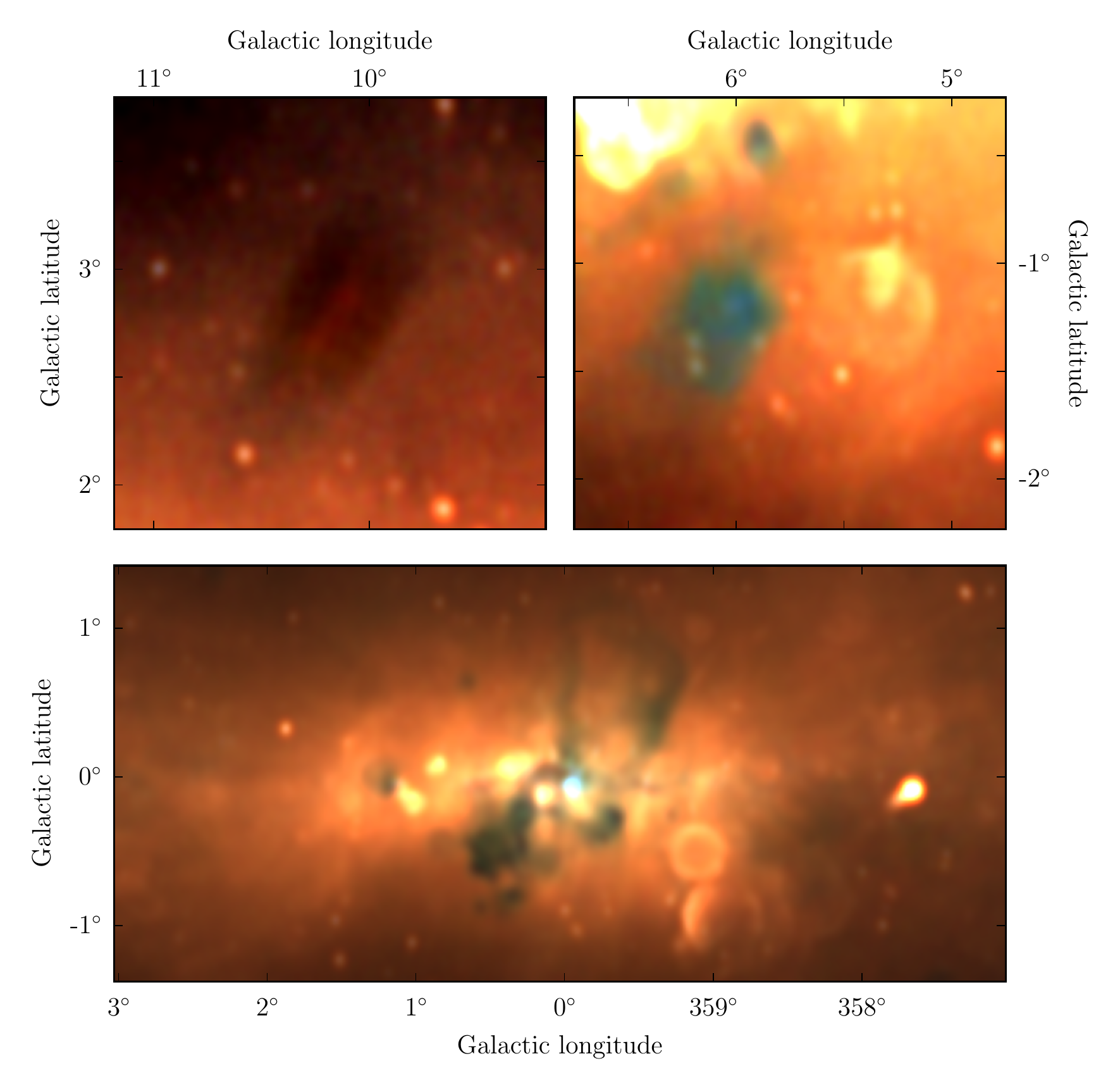}
    \caption{Three example images from this Galactic plane data release, using an RGB cube formed of the 72--103\,MHz (R), 103--134\,MHz (G), and 139--170\,MHz (B) data. The top left panel shows the Moon averaged over approximately one hour of observing; the top right panel shows on a known \textsc{Hii} region (left) and a known SNR (right); the bottom panel shows a view of the Galactic Centre. The colour ranges used are $-0.1$--4, $-0.1$--5, and $-0.5$--20\,Jy\perbeam~ for each panel, respectively.}
    \label{fig:thumbnails}
\end{figure*}

As visible in \Figs~\ref{fig:headline_week4} and~\ref{fig:headline_week2}, these images contain a wealth of data on Galactic objects. \Fig~\ref{fig:thumbnails} shows the discriminating power of the wide bandwidth in examining the physical origin of different kinds of emission. In the case of the Moon (top left panel), the disc obscures the background Galactic synchrotron emission, and reflects the FM radio emitted by the Earth, giving the centre a red appearance. The top right panel shows the known \textsc{Hii} region G$6.165-1.168$ \citep{1989ApJS...71..469L} and the known SNR ``Milne 56''\citep{1975AuJPA..37...75C};  free-free absorption and thermal emission gives the \textsc{Hii} region a distinctive blue appearance (positive $\alpha$) while the SNR has a fairly flat spectral index of $-0.2$ and appears slightly orange against the red of the steep-spectrum diffuse Galactic synchrotron. The lowest panel of  \Fig~\ref{fig:thumbnails} shows the GLEAM view of the Galactic centre, with strong free-free absorption and an intriguing loop of absorption perpendicular to the Galactic plane (Anderson et al., in prep).

Some studies have already been published using these data: \cite{2017MNRAS.465.3163S} and \cite{2018MNRAS.479.4041S} used the low-frequency absorbing properties of \textsc{Hii} regions to measure the cosmic ray emissivity of the Galactic plane along those lines-of-sight, Su et al. (in prep) produce a catalogue of all \textsc{Hii} regions selected from this region; Maxted et al. (submitted) examine in detail the $\gamma$-ray and radio properties of the young SNR G\,$23.11+0.18$; Hurley-Walker et al. (submitted) characterise 19 SNR candidates from the literature, including discriminating \textsc{Hii} regions from SNR candidates; and Hurley-Walker et al. (submitted) discover 27~supernova remnants, six with pulsar associations.

There remain many interesting features in these data, and the field is open to compare them with other recently-published surveys of the Galactic plane, such as the gamma-ray survey of \cite{2018A&A...612A...1H}, and the wideband infrared survey of \cite{2010AJ....140.1868W}. All images are available via the GLEAM VO server\footnote{\href{http://gleam-vo.icrar.org/gleam\_postage/q/form}{http://gleam-vo.icrar.org/gleam\_postage/q/form}} and SkyView\footnote{\href{http://skyview.gsfc.nasa.gov}{http://skyview.gsfc.nasa.gov}}.

\section{Conclusions}\label{sec:conclusions}

This work makes available a further \survarea~deg$^2$ of the GLEAM survey, using multi-scale cleaning to better deconvolve large-scale Galactic structure. For the latitude ranges $345^\circ < l < 67^\circ$, $180^\circ < l < 240^\circ$ , we provide images covering $|b|<10^\circ$; for the latter longitude range we also provide a compact source catalogue, while for the longitude range toward the Galactic Centre, we provide a compact source catalogue over $1^\circ\leq|b|\leq10^\circ$; the catalogue consists of \nsrc sources in total.

\begin{acknowledgements}

We thank the anonymous referee for their comments, which improved the quality of this paper. This scientific work makes use of the Murchison Radio-astronomy Observatory, operated by CSIRO. We acknowledge the Wajarri Yamatji people as the traditional owners of the Observatory site. Support for the operation of the MWA is provided by the Australian Government (NCRIS), under a contract to Curtin University administered by Astronomy Australia Limited. We acknowledge the Pawsey Supercomputing Centre which is supported by the Western Australian and Australian Governments. The National Radio Astronomy Observatory is a facility of the National Science Foundation operated under cooperative agreement by Associated Universities, Inc. We acknowledge the work and support of the developers of the following following python packages: Astropy \citet{TheAstropyCollaboration2013}, Numpy \citep{vaderwalt_numpy_2011}, and Scipy \citep{Jones_scipy_2001}. We also made extensive use of the visualisation and analysis packages DS9\footnote{\href{ds9.si.edu}{http://ds9.si.edu/site/Home.html}} and Topcat \citep{Taylor_topcat_2005}. This work was compiled in the very useful free online \LaTeX{} editor Overleaf.

\end{acknowledgements}


\bibliographystyle{pasa-mnras}
\bibliography{refs}

\end{document}